\documentclass[3p]{elsarticle}
\usepackage[center,scriptsize]{subfigure}
\usepackage{graphicx}
\usepackage{psfrag}
\usepackage{color}
\usepackage{booktabs}

\newcommand{\CD}{$\textrm{C}_{\textrm{\footnotesize D}} \:$}
\newcommand{\CL}{$\textrm{C}_{\textrm{\footnotesize L}} \:$}
\newcommand{\PD}{$\textrm{P}_{\textrm{\footnotesize D}} \:$}
\newcommand{\PL}{$\textrm{P}_{\textrm{\footnotesize L}} \:$}

\newcommand{\CDxx}[1]{$\textrm{C}_{\textrm{\footnotesize D}} #1 \:$}

\newcommand{\PDxx}[1]{$\textrm{P}_{\textrm{\footnotesize D}} #1 \:$}
\newcommand{\PLxx}[1]{$\textrm{P}_{\textrm{\footnotesize L}} #1 \:$}

\begin{document}

\author[label1,label2,label3]{K.~Szarf}
\author[label1,label2,label3]{G.~Combe\corref{cor}}
\ead{gael.combe@ujf-grenoble.fr}
\author[label1,label2,label3]{P.~Villard}

\cortext[cor]{Corresponding author}
\address[label1]{UJF, Laboratoire 3SR}
\address[label2]{Grenoble INP, Laboratoire 3SR}
\address[label3]{CNRS UMR 5521 \\ 3SR, BP53, 38041 Grenoble Cedex 9, France}

\begin{abstract}
We have simulated a granular medium using the Discrete Element Method to study the influence of particle shape  on the behaviour of the whole assembly of 5,000 grains. By comparing two shape groups: polygons and clumps of discs, we observed the effects of grain shape variations on both macro and microscopic levels and investigate the causes.
\end{abstract}

\maketitle
\begin{figure}[h]
 \begin{center}
  \includegraphics[angle=90,width=0.49\textwidth]{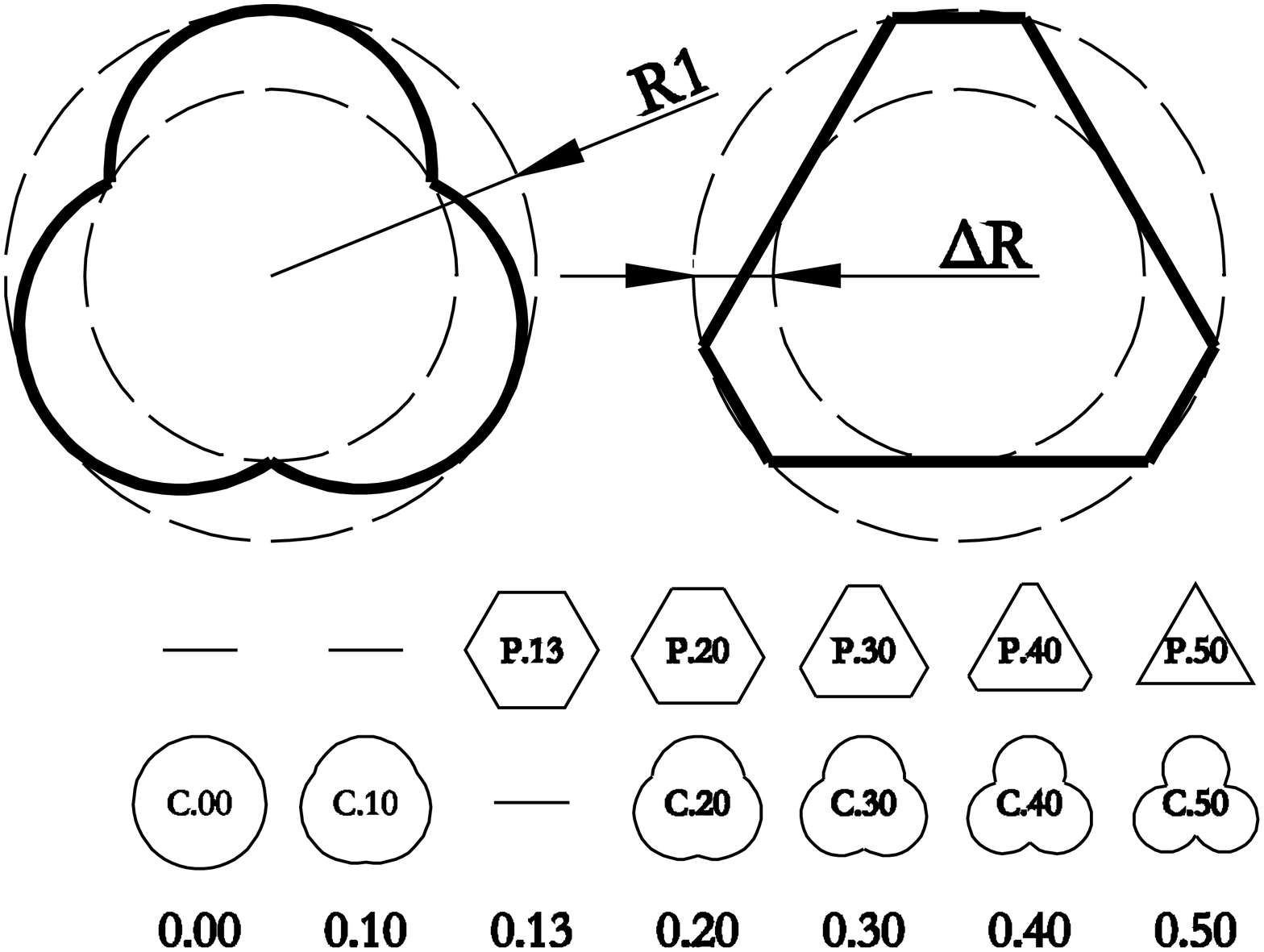}
 \end{center}
\end{figure}

\clearpage

\title{Polygons vs. clumps of discs: a numerical study of the influence of grain shape on the mechanical behaviour of granular materials}

\begin{abstract}
We performed a series of numerical vertical compression tests on assemblies of 2D granular material using a Discrete Element code and studied the results with regard to the grain shape. The samples consist of 5,000 grains made from either 3 overlapping discs (clumps - grains with concavities) or six-edged polygons (convex grains). These two grain type  have similar external envelopes, which is a function of a geometrical parameter $\alpha$.

In this paper, the numerical procedure applied is briefly presented followed by the description of the granular model used. Observations and mechanical analysis of dense and loose granular assemblies under isotropic loading are made. The mechanical response of our numerical granular samples is studied in the framework of the classical vertical compression test with constant lateral stress (biaxial test). The comparison of macroscopic responses of dense and loose samples with various grain shapes shows that when $\alpha$ is considered a concavity parameter, it is therefore a relevant variable for increasing mechanical performances of dense samples. When $\alpha$ is considered an envelope deviation from perfect sphericity, it can control mechanical performances for large strains. Finally, we present some remarks concerning the kinematics of the deformed samples: while some polygon samples subjected to a vertical compression present large damage zones (any polygon shape), dense samples made of clumps always exhibit thin reflecting shear bands.
\\
This paper was written as part of a CEGEO research project\footnote{www.granuloscience.com}

\end{abstract}


\maketitle


\section{Introduction}
A typical numerical approach to discrete element modeling of granular materials is to use simple shapes of particles (discs in 2D \cite{LJ00} or spheres in 3D \cite{THLA97}). Although the computation time is short using this method, these models cannot reflect some of the more complex aspects of real granular media behaviour, such as high shear resistance or high volumetric changes \cite{MGM99}. In order to model these mechanisms properly, physical phenomena (resistance to inter-granular rolling \cite{IWOD98,TOST02,GRC2007}) or other grain shapes (sphere aggregates \cite{SGV2009} or polyhedral grains \cite{ARPS2007}) must be used. The influence of grain shape is not yet fully understood. In this article, we will present our findings concerning the influence of grain shape (grain concavity in particular) on the mechanical behaviour of granular assemblies. We compared two groups of grains - convex irregular polygons and non-convex aggregates of three overlapping discs.

\section{Granular Model}
The granular model used consists of 5,000 polydisperse 2D frictional particles. Two kinds of grain shape are used: convex irregular polygons with six edges and non-convex particles made of aggregates of three overlapping discs called \emph{clumps}. These two shapes were chosen because of the similarity of their global contour (polygonal grains can be seen as a polygonal envelope of clumps made of three discs). As shown in Fig.~\ref{fig:shapes}, particle shape is defined by a parameter $\alpha= \frac{\Delta R}{R_1}$, where $R_1$ denotes the ex-circle radius of the particle and $\Delta R$ is the difference between the ex-circle and the in-circle radii, \cite{SCV2009,SCVDRS2009}. The in-circle must be fully contained in the particle. For non-convex clumps, $\alpha$ ranges from 0 (circle) to 0.5. For convex polygonal grains, $\alpha$ ranges from $1- \frac{\sqrt{3}}{2} \simeq  0.13$ (regular hexagons) to $0.5$ (equilateral triangles).  Some of the shapes used are presented at the bottom of Fig.~\ref{fig:shapes}. For each chosen $\alpha$, granular samples are made of polydisperse particles: the polydispersity of grains is determined by the radii of the grain ex-circle. In each sample, the chosen radii $R_1$ are such that the areas of the ex-circles are equally distributed between $S_m = \pi (R_m)^2$ and $S_M = \pi (R_M)^2 = \pi (3 R_m)^2$.

\section{Discrete Element Method}
Two-dimensional numerical simulations were carried out using the Discrete Element Method according to the principles of \emph{Molecular Dynamics} (MD) \cite{RAPA1995}. Two codes were used: PFC$^{2D}$ by ITASCA \cite{PFC1999} for clump simulation and a code capable of dealing with polygonal particles that was developed at the laboratory. Both codes use the same contact laws for contact forces computations \cite{CUNDS79}: grains interact in their contact points with a linear elastic law and Coulomb friction. The normal contact force $f_n$ is related to the normal interpenetration (or overlap) $h$ of the contact
\begin{equation}
\label{eq:fn}
 f_n = k_n \cdot h \quad ,
\end{equation}
as $f_n$ vanishes if contact disappears, i.e. $h= 0$. The tangential component $f_t$ of the contact force is proportional to the tangential elastic relative displacement, according to a stiffness coefficient $k_t$. The Coulomb condition $\vert f_t \vert~\le~\mu~f_n$ requires an incremental evaluation of $f_t$ in every time step, which leads to some amount of slip each time one of the equalities $f_t = \pm \mu f_n$ is imposed ($\mu$ corresponds to the contact friction coefficient). A normal viscous component as opposed to the relative normal motion of any pair of grains in contact is also added to the elastic force $f_n$. Such a term is often introduced to facilitate the mechanical equilibrium approach \cite{ACC2009}. In case of frictional assemblies under quasi-static loading, the influence of this viscous force (which is proportional to the normal relative velocity, using a damping coefficient $g_n$) is not significant \cite{JNRC2003} (elastic energy is mainly dissipated by Coulomb friction). Finally, the motion of grains is calculated by {solving Newton's equations using either a leap-frog (in PFC$^{2D}$) or third-order predictor-corrector discretisation scheme \cite{ALTI1987} (in the in-house software). This constitutes the only known difference between the two codes.

The principles of disc contact detection are well known \cite{CUNDS79,LUD1997}, and contact detection for clumps was solved in the same way: contact occurs at a point, the normal force value $f_n$ is calculated with eq. (\ref{eq:fn}) and its direction connects the centers of discs in contact, Fig.~\ref{fig:Contact_explanation}. Contact detection and contact force calculations between polygons do not use classical methods based on the area overlap between polygons \cite{HOPK1994,MAT98,AMLHV2005,AH2002}. The \emph{shadow overlap} technique proposed by J.J.~Moreau \cite{MORE2006}, which was originally applied within the \emph{Contact Dynamic} approach \cite{SCGDBM2006} for convex polygonal particles, was used. In our study this technique was adapted to the MD approach. Three  types of geometrical contact can exist between polygons: Corner-to-Corner, Corner-to-Edge \emph{(CE)} and Edge-to-Edge \emph{(EE)}. Corner-to-Corner contacts are geometrically (or mathematically) realistic but never occur in our simulations because of the numerical rounding errors. When dealing with \emph{(EE)} contact, contact detection involves two contact points and their associated overlaps $h$, Fig.~\ref{fig:contactCE_EE}. This is the main difference compared to the classical method (area overlap calculations) where only one contact is considered between the edges.

Finally, we may be interested in the main contact law parameters: the normal and tangential stiffness, $k_n$ and $k_t$, and the friction coefficient $\mu$. Assuming that samples would first be  loaded with a 2D isotropic stress $\sigma_0 = 10$~kN/m, the normal stiffness of contact $k_n$ was calculated according to the dimensionless 2D \emph{stiffness parameter} $\kappa = {k_n} / {\sigma_0}$ \cite{JNRC2003,CR2003,Gael,RC2005}. $\kappa$ expresses the mean level of contact deformation, $1 / \kappa = h / \left\langle 2R \right\rangle$, where $\left\langle R \right\rangle$ is the mean particle radius. In our simulations, $\kappa$ was arbitrarily set to $1,000$. As a comparison, a sample made of glass beams under isotropic loading of $100$~kPa reaches $\kappa = 3,000$. The tangential stiffness $k_t$ can be expressed as a fraction of the normal stiffness, $\tilde{k} = k_t/k_n$, $\tilde{k} > 0$. $\tilde{k} > 1$ may exhibit specific behaviour where \emph{Poisson coefficient} of grain assemblies become negative \cite{ECM1996,CDES1995,BT1988a,BT1988b}. Running several numerical simulations with various $\tilde{k}$, $0 < \tilde{k} \le 1$, \cite{Gael} have shown that if $0.5 \le \tilde{k} \le 1$, the macroscopic behaviour remains similar. Thus we arbitrarily set $\tilde{k}$ to $1$.

\section{Sample Preparation - Isotropic Compression}
\label{sec:isotropic}

Granular samples of $5,000$ grains are prepared in three steps: preparations start with a random spatial distribution of particle position inside a square made of four rigid walls. Secondly, the particles expand slowly until $\sigma_0 = 0.5$~kN/m is reached.  Finally, samples are isotropically loaded by wall displacement up to $\sigma_0 = 10$~kN/m. To obtain samples with different compacities, we may use various values of the inter-granular friction coefficient $\mu$ during the preparation \cite{CV2005}. When $\mu$ is set to zero, samples isotropically loaded up to $\sigma_0 = 10$~kN/m are dense and the compacity is \emph{maximal}. When {a strictly positive} value of $\mu$ is used instead, samples become looser and compacity decreases. In our study dense samples were prepared with $\mu = 0$ and the loose ones with $\mu$ equal $0.5$. 16 different samples (4~dense,~4~loose made of clumps and 4~dense,~4~loose made of polygons) for each $\alpha$ value were prepared\footnote{All the analyses presented in this article were carried out on mean results calculated over 4 samples of each density and each $\alpha$. Associated Standard Deviations will always be given, even if they are too small to be significant}. Dense samples will be written as \CD or \PD respectively for Clumps and Polygons. Loose samples will be denoted \CL or \PL. A subscript can be added. It then corresponds to the decimal part of the shape number $\alpha$. As an example, fragments of two dense samples with $\alpha = 0.30$, \CDxx{.30} and \PDxx{.30}, are displayed in figure \ref{fig:samples}.

For both clumps and polygons, contact between particles can occur at more than one contact point. There are four contact possibilities for clumps: single contact Fig.~\ref{fig:contactCE_EE}(i), double contact involving three discs Fig.~\ref{fig:contactCE_EE}(ii), double contact involving four discs Fig.\ref{fig:contactCE_EE}(iii) and triple contact Fig.~\ref{fig:contactCE_EE}(iv). By analogy with polygon contacts (Edge-to-Edge or Corner-to-Edge, Fig.\ref{fig:contactCE_EE}), all these contacts between clumps can be merged into two groups, \emph{(CE)} and \emph{(EE)}. In the \emph{(CE)} group, grains involved in a single contact (i) may rotate without sliding. Double contact (ii) allows rotation with sliding and eventually friction. Rotation and sliding of polygons meeting at a single contact point are not correlated. Group \emph{(EE)} contacts block the rotation of the grains. In the case of rotation of grains, contacts of this type would be lost. Therefore, the shape of the grains can be regarded as macro roughness.

For samples subjected to isotropic loading, we focus on two internal parameters that mainly determine the mechanical behaviour: the compacity $\xi$ and the coordination number $z^\ast$. Figure \ref{fig:compacity} shows the evolution of $\xi$ with $\alpha$ for dense and for loose samples. For \CD samples, $\xi$ evolution is bell-shaped and maximum compacity is reached for $0.2 \le \alpha \le 0.3$.   Similar observations were made by \cite{SCV2009,SCVDRS2009}. This also seems to be the case for \CL samples although the amplitude of the bell-shaped curve appears to be lower. The geometrical origin of these results deals with the  grain shape (concavity and grains envelope) and imbrication and the interlocking between grains, but appears to be complex to establish. While we might think that when two grains are in contact with a single point of contact (contact \emph{(CE)-type (i)}, Fig.~\ref{fig:contactCE_EE}), this would tend to increase the local porosity and thus reduce the overall compacity, this does not seem to be the case: figure \ref{fig:contactCE} shows clearly that for \CD samples, the proportion of contact \emph{(CE)-type (i)} is at its maximum for $0.2 \le \alpha \le 0.3$ in the case of dense samples and decreases linearly with $\alpha$ in the case of \CL samples. This completely goes against the $\xi$ trend, even if it is clearly established that when $\xi$ is high (\CD samples), the proportion of \emph{(CE)-type (i)}  contacts is lower than in the case where the compacity is low (\CL samples).

For samples made of polygons, the dependence of $\xi$ on $\alpha$ is not clear either. For \PD samples, $\xi$ is almost constant for $0.2 \le \alpha \le 0.3$. If we exclude the \PDxx{.40} sample, which behaves in a peculiar way, we can notice that for samples made of grains with regular shapes, corresponding to $\alpha = 0.13$ and $\alpha = 0.5$, the compacity $\xi$ is smaller than for samples made of grains which have shape irregularities ($0.13 < \alpha < 0.5$). The relationship between $\xi$ and the percentage of \emph{(CE)-type (i)} contact is again impossible to establish. Furthermore, it is worth observing that when $0.2 \le \alpha \le 0.3$, \CD and \PD samples show very similar compacity. Here, grain envelopes seems to be a cleverer interpretation of $\alpha$ parameter than grain imbrication, which is only relevant for clumps. Finally, we should note that the angle of friction used during the preparation does not seem to have a major influence on the compacity of samples made of polygons when $0.2 < \alpha < 0.3$.

For samples subjected to isotropic loading, we studied the \emph{coordination number} $z^{\ast}$ corresponding to the mean number of contacts per grain. Here only grains that have two or more compression forces, and therefore take part in the load transfer, were considered. For samples made of frictionless perfectly rigid discs, $z^{\ast} = 4$ \cite{AR2007}. Because $\kappa$ is not infinite  in our study, in the samples made of frictionless circular particles (\CDxx{.00}), $z^{\ast} = 4.093 \pm 0.005$ is greater but still very close to the reference value. $z^{\ast}$ is evaluated for clumps and polygons and both dense and loose samples. The dependence of $z^{\ast}$ according to $\alpha$ is shown in Fig.~\ref{fig:coordinationnumber}. For \CL samples, $z^{\ast}$ increases linearly with $\alpha$, like \PL samples, but for \CL, it can be directly correlated to the percentage of \emph{(CE)-type (i)} contacts which decrease with $\alpha$, and then increase $z^{\ast}$. For dense samples, the percentage of \emph{(CE)-type (i)} contacts did not vary too much. $z^{\ast}$ is constant for \CD and \PD samples.

\section{Macromechanical Response of Granular Assembly Loaded in Vertical Compression Test}
\label{sec:macro}

The samples were tested in a 2D strain controlled vertical compression test, also called \emph{biaxial test}. Vertical stress $\sigma_1$ was applied by increasing the compressive vertical strain $\varepsilon_1$ while lateral $\sigma_3$ remained constant. The loading velocities were chosen according to the dimensionless \emph{inertial number} $I=\dot{\varepsilon}_1\sqrt{\frac{\left\langle m\right\rangle}{\sigma_3}}$ \cite{RC2005} where $\dot{\varepsilon}_1$ denotes the strain rate and $\left\langle m\right\rangle$ the typical mass of a grain. It describes the level of dynamic effects in the sample. For quasi-static evolutions, the value of $I$ should be low. $I$ value was set to $5\cdot10^{-5}$ for clumps and for polygons samples, regardless of the code used. During the vertical compression in both dense and loose samples, the same value of contact friction coefficient $\mu = 0.5$ was used. The mechanical responses of the samples are plotted on $\eta$ vs. $\varepsilon_1$ charts and shown in figure \ref{fig:macroscopicCurves}. $\eta = {t}/{s}$, $t={(\sigma_1 - \sigma_3)}/{2}$ is half of the deviator stress and $s=(\sigma_1 + \sigma_3)/{2}$ is the mean stress. Extracted from $\eta$-$\varepsilon_1$ curves shown in figure \ref{fig:macroscopicCurves}, friction angles at the peak $\phi_p$ and at the threshold $\phi_t$ are given figure \ref{fig:FrictionAngle}. Average dilatancy angles extracted from Fig.~\ref{fig:macroscopicCurves-vol} are presented in the figure \ref{fig:dillatation_angle}.

For dense samples made of clumps, \CD, we can observe in figure \ref{fig:Cd} that the macroscopic shear resistance increases with $\alpha$. Although \CDxx{.10} implies grains with a small $\alpha$, the mechanical response of the sample exhibits remarkable increase of the maximum deviator in comparison to disc samples \CDxx{.00} where rotations of particles are not potentially disturbed by the grain shape. For \CL samples, there is no peak value of the friction angle, $\phi_{p} = \phi_{t}$, figure \ref{fig:Cl}. Thus, for all clump samples, both peak $\phi_{p}$ and threshold $\phi_{t}$ friction angle values increase along with $\alpha$. $\phi_{t}$ increases proportionally with $\alpha$ while the increase of $\phi_{p}$ is nonlinear and seems to be asymptotic for $\alpha \ge 0.4$.

For \PD samples, $\phi_{p}$ values slightly decrease linearly along $\alpha$, while $\phi_{t}$ values increases, Fig.~\ref{fig:FrictionAngle}.

For \PL samples, we can notice in figure \ref{fig:Pl} that the macroscopic curves show typical behaviour similar to that of loose samples when $\alpha < 0.3$ and a typical behaviour characteristic of dense samples for $\alpha > 0.3$. This kind of behaviour deals with the values of the initial compacity of the samples shown in figure \ref{fig:compacity} where we can observe that $\xi$ values for \PD and \PL samples are very close when $\alpha \leq 0.3$. If we focus only on $\phi_t$ for \PL samples, we can observe (Fig.~\ref{fig:FrictionAngle}) an increase of the friction angle with $\alpha$, except for samples made of triangles, $\alpha = 0.5$, which always behave in a specific way\footnote{Note that triangle is the only shape with 3 edges}. In conclusion, it can be noted that adding some shape irregularity by increasing $\alpha$ always leads to an increase of the macroscopic angle of friction in the critical state. This is the case for clumps and polygons with a constant microscopic friction angle $\mu$.

This influence of grain geometry is in line with a previous study by Salot et al. \cite{SGV2009}. Lastly, as we can see in figure \ref{fig:macroscopicCurves}, $\alpha$ does not explicitly influence Young's modulus. $E$ is linked to the rigidity matrix and therefore to $z^{\ast}$ \cite{ARIII2007}, which is constant for dense samples (Fig.~\ref{fig:coordinationnumber}).

Similarly, particle concavity does not particularly influence   the average dilatancy angle $\psi$ values ($\sin \psi = \frac{d \varepsilon_1 + d \varepsilon_3}{d \varepsilon_1 - d \varepsilon_3}$) of dense and loose clump samples (Fig.~\ref{fig:dillatation_angle}). On the other hand, $\psi$ is lower for polygons with higher values of $\alpha$ (closer to triangular shape) than for those more similar to hexagons. 

In figure \ref{fig:macroscopicCurves-vol}, volumetric changes in some samples are illustrated. For both dense clump and polygon samples, Fig.~\ref{fig:Cd-vol} and \ref{fig:Pd-vol}, the volume\footnote{For convenience we resort to the vocabulary pertaining to 3D triaxial tests} increases mainly during vertical compression, after a small contraction due to the stiffness of the contacts \cite{JNRC2003}. The volumetric increase for \CD samples is quite similar from one $\alpha$ to another (Fig.~\ref{fig:Cd-vol}). On the other hand, $\alpha$ clearly influences the volumetric change of \PD samples but this influence seems erratic. Nevertheless, \PD  \phantom{0} samples show greater total dilatancy than \CD  samples. \CL  samples (Fig.~\ref{fig:Cl-vol}) behave like loose sands and contract all throughout the compression test. It is more complex for \PL samples (Fig.~\ref{fig:Pl-vol}) for which the setting-up process remains problematic for some values of $\alpha$. 

\section{Micromechanical Analysis}

From the macroscopic results exposed in the previous section, two main observations can be established: for dense samples made of clumps, the evolution of the angle of friction at the stress peak $\phi_p$ varies significantly with $\alpha$. The geometrical imbrication between grains in contact, which depend on $\alpha$, may be one of the micro-mechanical origins of these results.
Secondly, for all samples, dense or loose, made of clumps or polygons, it was established that the angle of friction $\phi_t$ at the end of biaxial tests increases with $\alpha$ and is independent of the initial state. This result is proof of the role of the grain envelope in the mechanical behaviour rather than some inter-granular imbrication considerations. 
In this section we will try to gather evidence for these proposals.

\subsection{Contact proportion evolution for dense samples made of clumps}

Single contacts are mainly involved in all the grains samples tested, Fig.~\ref{fig:contactCE} (\emph{(CE)} contacts for polygons or \emph{(CE)-type (i)} contacts for clumps -- Fig.~\ref{fig:contactCE_EE}). During loading, it can be noted that the percentage of single contacts increases. Because dense samples were prepared with no friction, compacity of each assemblies of grains subjected to isotropic loading is constant and always maximal. As a consequence, even if samples can exhibit slight contractancy (related to dimensionless contact stiffness $\kappa$, \cite{RC2002}), the total number of contacts in each sample during a vertical compression systematically decreases. Therefore, in order to compare different contact type proportions in different phases of the test, we suggest balancing the decrease in the total number of contacts using the coefficient $\omega^{*} = {N^{*}_{\varepsilon_b}}/{N^{*}_{\varepsilon_a}}$. $N^{*}_{\varepsilon_a}$ and $N^{*}_{\varepsilon_b}$ represent the total number of neighbouring contacts\footnote{When two grains are in contact via 1, 2 or 3 contact points, only one contact is counted} at the given vertical strains $\varepsilon_1$ ($\varepsilon_a$ or $\varepsilon_b$) in the samples. Here, we suggest focusing on the evolution of two new contact groups for clumps:
\begin{itemize}
\item \emph{(SC)} single contact between grains, known as \emph{simple} contact,
\item \emph{(CC)} multiple contacts between grains, hereafter called \emph{complex} contacts.
\end{itemize}

These two new groups can be examined in figure \ref{fig:contact2}.

We observed the evolution of clump contact numbers of each group between two successive stages by normalising this evolution with $\omega^{*}$. We thus defined a new variable $\lambda =  \omega^{*} \cdot {N_{\varepsilon_a}}/{N_{\varepsilon_b}}$, where $N_{\varepsilon_a}$ and $N_{\varepsilon_b}$ denote the number of \emph{(SC)} or \emph{(CC)} on two different levels ${\varepsilon_a}$ and ${\varepsilon_b}$ of the vertical strain $\varepsilon_1$. In figure \ref{fig:4contacts_peak-iso}, $\lambda$ evolution is calculated between $\varepsilon_a = {\varepsilon_i}$ (isotropic state) and $\varepsilon_b = {\varepsilon_p}$ (maximum stress deviator). We can observe that for $\alpha = 0.10$, $\lambda$ is smaller than 1 for \emph{(CC)} and greater than 1 for \emph{(SC)}. This can be regarded as a transformation of \emph{(CC)} into \emph{(SC)} between these two stages. If all the complex contacts transformed into simple, graphic points were  at an equal distance from 1, $1 - \lambda_{\textrm{\small\emph{(CC)}}} = \lambda_{\textrm{\small\emph{(SC)}}} - 1$. If $1 - \lambda_{\textrm{\small\emph{(CC)}}} > \lambda_{\textrm{\small\emph{(SC)}}} - 1$, it means that some complex contacts transform into simple contacts but some of them also disappear.

When $\alpha$ goes to $0.5$, these transformations are still active but with less intensity. Geometrical imbrications between clumps increase with $\alpha$ and are ``more difficult to lose" during the biaxial tests. It is also interesting to observe that $\lambda$ seems to reach a threshold when $\alpha \ge 0.4$, Fig.~\ref{fig:4contacts_peak-iso}. This last observation can be correlated to the evolution of $\phi_p$, which also reaches a threshold for the same value of $\alpha$, Fig.~\ref{fig:FrictionAngle}.

Focusing on $\lambda$ between the peak and the critical state, $\varepsilon_a = \varepsilon_p$ and $\varepsilon_b = \varepsilon_c$, Fig.~\ref{fig:4contacts_peak-crit}, we can observe that the increase of simple contacts is small for every $\alpha$ ($\lambda_{\textrm{\small\emph{(SC)}}} - 1 \le 0.1$) and complex contacts are mainly lost $1 - \lambda_{\textrm{\small\emph{(CC)}}} > 0.1$, especially when $\alpha$ is small. The greater  $\alpha$ is, the smaller the proportion of complex contacts lost (grain imbrications are destroyed less). This may be a clue that $\phi_t$ of clumps increases with $\alpha$, as seen in Fig.~\ref{fig:FrictionAngle}. Nevertheless, some new investigations on the evolution of contact orientations are proposed in the next section.

\subsection{Evolution of contact fabric for \CD samples}

Contact direction and its evolution during the vertical compression tests are often analysed \cite{PandG1997}. Focusing on the first part of the mechanical behaviour, from the isotropic state to the stress peak for example, it is well known that in dense samples, contacts are mainly lost in the extension direction and gained in the direction of compression, \cite{CCL97}. In figures \ref{fig:OrientAllContact-a} to \ref{fig:OrientAllContact-d}, for two values of $\alpha$, we present statistical analysis of the evolution of contact direction by the evaluation of $\mathcal{P}(\theta) = N_{\varepsilon_b}(\theta) / N_{\varepsilon_a}(\theta)$, where ${\varepsilon_a}$ and ${\varepsilon_b}$ correspond to two successive vertical strains levels and where $N_{\varepsilon_x}(\theta)$ is the number of contacts in the direction $\theta$. $\mathcal{P}(\theta) = 1$ expresses that the number of contacts in the direction $\theta$ remains constant between the two configurations studied. If $\mathcal{P}(\theta) < 1$, contacts are lost and if $\mathcal{P}(\theta) > 1$, contacts are gained in the direction $\theta$. Integrated over $\theta$, $\langle \mathcal{P} \rangle$ is a global evaluation of the proportion of gained or lost contacts.

We focus on the evolution of contact anisotropy between the isotropic state and the peak, Fig.~\ref{fig:OrientAllContact-a} for $\alpha = 0.2$ and Fig.~\ref{fig:OrientAllContact-c} for $\alpha = 0.5$, we can observe that there is no contact gain in any direction: in the compression direction, the number of contacts remains constant $\mathcal{P}(\theta) \simeq 1$ and the number of contacts decreases in the extension direction $\mathcal{P}(\theta) < 1$. The mean value of $\mathcal{P}$ over $\theta$ is smaller than 1 for both $\alpha$ discussed here and also for the other $\alpha$ studied. During the vertical compression of \CD, sample contacts are mainly lost. Finally, we noticed that the greater $\alpha$ is, the smaller the amount of contacts lost in the extension direction. By analysing the contacts change in direction between the peak and the critical state, $\varepsilon_p$ to $\varepsilon_c$, Fig.~\ref{fig:OrientAllContact-b} and \ref{fig:OrientAllContact-d}, opposite tendencies emerge: contacts are mainly lost in the compression direction and gained in the extension direction, Fig.~\ref{fig:OrientAllContact-b}, with a less intensive effect for $\alpha = 0.5$,  Fig.~\ref{fig:OrientAllContact-d}. At this stage of analysis, we are not yet able to distinguish the nature of the contacts involved in these observations. An analysis of clustered contacts, as outlined below, is thus necessary.

We now suggest the same contact direction analysis but for $\emph(SC)$ and $\emph(CC)$ groups (simple and complex contacts). Statistical analysis of the evolution of contact orientation from the isotropic state $\varepsilon_i$ to the peak $\varepsilon_p$ for $\emph(SC)$ and $\emph(CC)$ groups is shown in figures \ref{fig:OrientAContact-IsoPeak-a} to \ref{fig:OrientAContact-IsoPeak-d}. On one hand, Fig.~\ref{fig:OrientAContact-IsoPeak-a} and \ref{fig:OrientAContact-IsoPeak-b} show that  in the compression direction \emph{(SC)} contacts are gained ($\alpha = 0.2$) or are kept ($\alpha=0.5$). \emph{(SC)} contacts are mainly lost in the extension direction, with a more pronounced amplitude when $\alpha$ is small. On the other hand, we can observe that \emph{(CC)}, Figs.~\ref{fig:OrientAContact-IsoPeak-c} and \ref{fig:OrientAContact-IsoPeak-d}, are lost in every direction with some variations depending on $\theta$. Nevertheless, complex contacts are more persistent when $\alpha$ is greater ($\langle \mathcal{P} \rangle$ is greater for $\alpha = 0.5$ because of grain imbrications).

The statistical analysis of the evolution of contact direction between the peak $\varepsilon_p$ and the critical state $\varepsilon_c$ shown in the figures~\ref{fig:OrientAContact-PeakCrit-a} to \ref{fig:OrientAContact-PeakCrit-d} confirms the tendency shown in figures \ref{fig:OrientAllContact-b} and \ref{fig:OrientAllContact-d}: simple and complex contacts are lost in the compression direction. For \emph{(SC)} with $\alpha = 0.2$, $\langle \mathcal{P} \rangle = 1$, Fig.~\ref{fig:OrientAContact-PeakCrit-a}: although the number of $\emph{(SC)}$ decreases in the compression direction and increases in the extension direction, the number of simple contacts remains constant during the mechanical test from the peak to the critical state.  When $\alpha = 0.5$, the number of simple contacts decreases $\langle \mathcal{P} \rangle = 0.9$, especially in the compression direction. Complex contacts are the ones that are lost the most ($\langle \mathcal{P} \rangle$ is always lower than 1, regardless of the value of $\alpha$). Nevertheless, it is desirable to make a distinction based on $\alpha$: the amount of \emph{(CC)} lost is smaller for $\alpha = 0.5$ ($\langle \mathcal{P} \rangle = 0.8$, Fig.~\ref{fig:OrientAContact-PeakCrit-c}) than for $\alpha = 0.2$ ($\langle \mathcal{P} \rangle = 0.7$, Fig.~\ref{fig:OrientAContact-PeakCrit-d}). The proportion of \emph{(CC)} lost in the compression direction is bigger for $\alpha = 0.2$, like if vertical contact chains were more unstable or less persistent when $\alpha$ tends to $0$. 

Further studies, taking into account the intensity of contact forces and their propensity to be stronger in the case of complex contacts \cite{ARP2007} would provide more certainty about possible links between the observations made above and improvement of the mechanical property $\phi_t$ measured at $\varepsilon_c$, Fig.\ref{fig:FrictionAngle}.

\subsection{Evolution of comparative contact proportions in compression tests for clumps and polygons}

Focusing on the critical phase only, contact observations are now based on the division into two contacts groups corner-to-edge \emph{(CE)} and edge-to-edge \emph{(EE)} (Fig.~\ref{fig:contactCE_EE}). The evolution of \emph{(CE)} contacts according to  $\alpha$  is shown in figure \ref{fig:corner-edge}. \emph{(EE)} contacts can be easily deduced by subtracting \emph{(CE)} contacts percentage from 100\%. On one hand, the percentage of \emph{(CE)} contacts  does not depend on the initial compacity of the sample; dense and loose samples exhibit a similar trend. On the other hand, \emph{(CE)} contact percentages for  \CD and \CL samples decrease linearly with $\alpha$. A different tendency is observed for \PD and \PL  samples where the contact percentage remains more or less the same except for $\alpha = 0.5$. We can even observe that when $\alpha = 0.5$, \emph{(CE)} contacts and \emph{(EE)} contact percentages are close: for this special value of $\alpha$, clump and polygon envelopes converge.

\subsection{Influence of shape on local strain analysis}

We focused on the strain localisation in the samples in order to study the macroscopic rupture and its origin. Two approaches were used: \emph{local strain maps} and \emph{shear localisation indicator} $S_2$ \cite{SDS1993}. By comparing particle kinematics in the isotropic state $\varepsilon_1 = \varepsilon_i = 0$ and in the deformed stage $\varepsilon_1$, we calculated local strains using Delauney triangulation as in \cite{CCL97} (Delauney triangle corners correspond to the mass centers of particles). Using the second strain invariant, $I_{\varepsilon_d} = \varepsilon_I -\varepsilon_{II}$, where $\varepsilon_I$ and $\varepsilon_{II}$ are respectively the major and the minor principal strains, we illustrate the shear localisation in figure \ref{fig:shearmaps_poly_vs_clumps}. We should notice that such shear localisation patterns, also called shear bands, are often observed on granular materials confined by more or less rigid boundaries, or even numerical or experimental considerations \cite{CCL97, LANIER1976, LJ2000, VD2004}. Even if the number of shear bands depends on the macroscopic strain levels applied to samples, different patterns (multiple shear zones) seem to exist when periodic boundary conditions are used \cite{KB2009}.

The shear localisation indicator $S_2$ is defined as 

\begin{equation}
 S_2 = \frac{1}{N_t}\frac{\left(\displaystyle \sum_{i=1}^{N_t}I_{\varepsilon_d}\right)^2}{\displaystyle \sum_{i=1}^{N_t}I_{\varepsilon_d}^2} \quad ,
\end{equation}

\noindent where $I_{\varepsilon_d}$ is the second invariant of the strain tensor and $N_t$ is the total number of Delaunay triangles. In a sense, value of $S_2$ can be regarded as a percentage of a distorted sample area.

Figure \ref{fig:shear_indicator} gives the evolution of $S_2$ according to  $\varepsilon_1$ for several dense and loose samples made of clumps and polygons. Regardless of the sample studied, the sheared area always reaches a maximum which is at least greater than $50\%$. From evolution of $S_2$, we may encounter two types of behaviour: samples for which $S_2$ reaches a maximum and then decreases and stabilises, and a second type where $S_2$ increases asymptotically towards a maximum.

We have observed that when $S_2$ reaches a maximum and later decreases to reach a threshold value, it always corresponds to samples which were identified as dense samples because $\phi_p > \phi_t$ (for example \CDxx{.50}, \PDxx{.24}, \PLxx{.28} samples of the figure \ref{fig:shearmaps_poly_vs_clumps}). On the contrary, when $\phi_p \sim \phi_t$, samples can be classified as loose samples. In this case, $S_2$ continuously increases from $\varepsilon_1 = 0$ to $10\%$ to eventually reach a threshold close to $S_2 \simeq 60\%$.

In the case of dense samples, the asymptotic value of $S_2$ is interesting because it clearly shows higher values for \PD samples than for \CD samples. Coupling this quantitative result with the qualitative observation of shear maps like in the figure \ref{fig:shearmaps_poly_vs_clumps}, it is obvious that localised zones in \PD samples are wider than in \CD samples. This result is consistent with the overall dilatancy of samples: \PD samples globally expand more than \CD ones. For loose samples, the sheared area always corresponds to approximately $60\%$ of the samples, regardless of the grain shape.

\section{Conclusions and discussion}

The aim of this article was to present some new investigations on the mechanical influences of particle shape in granular assemblies in the framework of numerical simulations performed with Discrete Element Method. First, a grain geometry parameter $\alpha$ was defined by the CEGEO research team. For particles called clumps, made of 3 overlapping discs, $\alpha$ is a measure of the grain concavity. 6-edged convex polygonal grains were also ruled by $\alpha$. The overall envelope depending on $\alpha$ for each type of particles used in the studied granular model was the common feature. Our numerical simulations were performed with the Discrete Element Method adapted to each grain shape. For particles made of discs (clumps), the commercial code PFC$^{2D}$ by ITASCA was used. For polygonal particles, we developed our own computer code which implements special contact detection between objects in the framework of Molecular Dynamic approach. In this article, we highlighted that changing the grain geometry, influences granular assembly mechanical behaviour under the classical vertical compression test in 2D. More complex grain shapes allow higher levels of internal friction angle and large volumetric strains to be reached compared to simple discs. Some clear differences in the behaviour of polygon (convex) and clump (non-convex) assemblies were shown. We should also note that the particle shapes chosen also demonstrate similarities, caused by the global envelope, that justify the comparison. The generation and compaction of granular assemblies was presented. By using two extreme values of the inter-granular friction angle $\mu$, dense and loose samples were prepared, both for samples made of clumps and polygons.

Firstly, by focusing on the macroscopic mechanical behaviour of our granular model we show that loose samples composed of polygons with low values of $\alpha$ present behaviour typical to moderately dense granular samples. For these granular materials, the initial contracting stage was not only due to contact stiffness, but was also influenced by large inter-granular reorganisation. Apart from this, loose and dense samples of all shapes behaved as expected (loose samples only show contracting behaviour while dense ones mostly exhibit major dilatancy), showing similar behaviour when discussing friction residual angles $\phi_t$ or contact percentages. All samples show higher values of internal friction angles $\phi_p$ and $\phi_t$ than samples made of only circular grains where each particle is a disc. The correlations between shape parameter $\alpha$ and friction angles are different for clumps and polygons. On one hand, for dense clump samples $\phi_p$ increases with $\alpha$ and seems to reach an asymptotic value $\phi_p = 40^\circ$. On the other hand, $\phi_p$ linearly decreases when $\alpha$ shifts from $0.13$ to $0.5$. In this case, the particular case of the triangular shape ($\alpha = 0.5$) is also discussed briefly. The overall dilatancy of clump samples is greater than that of disc assemblies, but spectacularly smaller than dilatancy of polygons.

Secondly, on the granular scale, we suggested correlating macroscopic observations by means of contact evolution analysis, which led us to introduce several groups of contacts between particles. Thus, we observed that multiple contacts between clumps transform into simple contacts and that this process depends on the size of concavities, i.e. $\alpha$. We tend to associate this with an increase of shear resistance in the case of dense granular samples made of clumps. However, for a better understanding of the role of $\alpha$ in the magnitude of contact forces and their influence on the macroscopic repercussions, complementary studies need to be carried out.

Focusing on granular assembly failure, we studied the localisation of shear bands and tried to characterise it by a scalar. It was observed that reflecting shear bands were thinner in dense samples made of clumps than in those made of polygons, regardless of the $\alpha$ type, thus highlighting  the evident effects of the geometrical imbrications of clumps. Polygon samples gradually create wide shear bands, while for samples made of clumps, the appearance of shear bands are more immediate.

Although the meaning and implications of the parameters $\alpha$ have been presented in this article, it needs to be clarified further. Nevertheless, from the comparisons between polygons and clumps, two trends seem to emerge: for very dense samples made of clumps, a large $\alpha$ naturally implies imbrications between particles. $\alpha$ is thus a measure of clump concavity. With the simulations exposed in this article, we can deduce that the larger the concavities are, the higher the angle of friction at the peak $\phi_p$ is. 

For the samples made of polygons, this does not apply. Indeed, there are no imbrications between the grains and $\alpha$ is thus a measure of grain sphericity. We have shown that in dense samples made of polygons, the percentage of a single contact in the isotropic state increases with $\alpha$. In corollary, we can observe that the angles of friction peak decrease slightly with $\alpha$. 

For loose samples \PL or \CL, or for \PD and \CD samples close to the critical state, $\phi_t$ is the parameter which characterises the failure. We have shown that for large strains, contacts between grains are mostly single. Therefore, clump imbrications are less involved in the evolution of $\phi_t$. Furthermore, $\phi_t$, which increases linearly with $\alpha$, increases with the same rate regardless of the grain shape. $\alpha$ should thus be only regarded as a parameter of spherical grains.

\section*{Acknowledgments}

This study was carried out as part of a CEGEO\footnote{www.granuloscience.com} research project. The authors would like to express special thanks to F.~Radjai, C.~Nouguier for fruitful discussions and to F.~Nicot for his very useful suggestions. The authors are indebted to J.J.~Moreau for his guidance on algorithm for contact detection between polygon objects.   

\bibliographystyle{elsarticle-num}
\bibliography{PowderT-GCOMBE-bib}

\end{document}